\documentclass[aps,preprint,amsmath,amssymb,amsfonts,nofootinbib]{revtex4-1}
\usepackage{epsfig}
\usepackage{graphicx}
\usepackage{dcolumn}
\usepackage{bm}
\usepackage{amsthm}
\usepackage{amsmath}
\usepackage{color}
\newcommand{\beq}{\begin{equation}}
\newcommand{\eeq}{\end{equation}}
\newcommand{\bea}{\begin{eqnarray}}
\newcommand{\eea}{\end{eqnarray}}

\begin{document}

\title{Holography and $AdS_2$ gravity with a dynamical aether}
\author{Christopher Eling}
\email{cteling@gmail.com}
\affiliation{Rudolf Peierls Centre for Theoretical Physics, University of Oxford, 1 Keble Road, Oxford OX1 3NP, UK}

\begin{abstract}
We study two-dimensional  Einstein-aether (or equivalently Ho\v{r}ava-Lifshitz) gravity, which has an $AdS_2$ solution. We examine various properties of this solution in the context of holography. We first show that the asymptotic symmetry group is the full set of time reparametrizations, the one-dimensional conformal group. At the same time there are configurations with finite energy and temperature, which indicate a violation of the Ward identity associated with one-dimensional conformal invariance. These solutions are characterized by a universal causal horizon and we show that the associated entropy of the universal horizon scales with the logarithm of the temperature. We discuss the puzzles associated with this result and argue that the violation of the Ward identity is associated with a type of explicit breaking of time reparametrizations in the hypothetical $0+1$ dimensional dual system.

\end{abstract}

\maketitle

\section{Introduction}

One of the most mysterious aspects of gauge/gravity dualities is a hypothetical $AdS_2$/$CFT_1$ correspondence mapping a two-dimensional gravity theory in anti- de Sitter spacetime into a one-dimensional conformal theory. This type of duality is of interest since generically the geometry in the near horizon limit of extremal black holes contains an $AdS_2$ factor. On the non-gravitational side of the duality the term ``one-dimensional" refers to a $0+1$ dimensional quantum mechanical model. Conformal means an invariance under arbitrary reparametrizations of time $t \rightarrow f(t)$, which are the asymptotic symmetries of $AdS_2$ spacetime \cite{Hotta:1998iq,Cadoni:1999ja,NavarroSalas:1999up}. While physics in lower dimensions is simpler, in this case it appears to be too simple. One way of stating the problem is that the conformal Ward identity, which generically enforces the tracelessness of the stress-tensor, implies in one-dimension that the energy is zero since the stress tensor only has one component $T^{tt}$. Thus the quantum mechanical theory has no dynamics. On the gravity side this reflected in the fact that pure Einstein gravity is trivial and the Einstein-Hilbert action in two dimensions is a topological term.

An early model of a non-trivial two dimensional theory of quantum gravity was studied by Polyakov \cite{Polyakov:1981rd}. The action is given by the non-local Polyakov term which generates the trace anomaly in two-dimensions. In the conformal gauge this reduces to Liouville gravity, which is a special case of a general class of dilaton gravity theories. These theories have solutions with an $AdS_2$ metric plus a non-trivial dilaton in the bulk.  Recently,  these results have been interpreted in terms of a nearly $AdS_2$/$CFT_1$ ($NAdS_2$/$NCFT_1$) correspondence  (see, e.g. \cite{Almheiri:2014cka,Jensen:2016pah,Maldacena:2016upp,Engelsoy:2016xyb,Mandal:2017thl}). On the gravity side, the dilaton explicitly breaks the time reparametrization asymptotic symmetry.  The dual description is in terms of the first finite temperature/energy corrections away from the infrared to a type of quantum mechanical model of interacting fermions, the Sachdev-Ye-Kitaev model. This model has has an emergent reparametrization invariance in the $T \rightarrow 0$ limit \cite{Kitaev,Maldacena:2016hyu}.

In this paper we will consider another theory of gravity in two dimensions, Einstein-aether theory \cite{Eling:2006xg}. In this theory the metric is coupled to a dynamical unit timelike vector field, the ``aether".  One can also recast the theory into a Ho\v{r}ava-Lifshitz form\footnote{Specifically the non-projectable, extended version of the theory without an additional $U(1)$ symmetry \cite{Blas:2009qj,Jacobson:2010mx}}, which has a preferred foliation of time and therefore is invariant only under foliation preserving diffeomorphisms \cite{Horava:2009uw}.  In higher dimensions this class of theories has been studied as a potential holographic dual to strongly coupled non-relativistic field theories invariant under Lifshitz scaling symmetries, where Lorentz invariance is broken, see e.g. \cite{Janiszewski:2012nb, Griffin:2012qx, Eling:2014saa,Davison:2016auk}. In Section II we discuss the properties of the theory in two-dimensions and show it has solution with an $AdS_2$ metric plus a non-trivial aether profile in the bulk.

As a first probe of holographic properties, we investigate the asymptotic symmetries of this solution in Section III. We find that the time reparametrizations of $AdS_2$ are unbroken by the aether. Despite this fact, we show that there are configurations with finite energy and temperature. In the usual case of a higher dimensional CFT, the presence of a finite temperature introduces a scale which spontaneously breaks the conformal symmetry. One finds an energy associated with the thermal state, but the Ward identity enforcing the traceless stress tensor still holds. However, in $0+1$ dimensions there is a conflict between spontaneous breaking and the Ward identity. A finite temperature configuration has finite energy, but this is inconsistent with required vanishing $T^{tt}$ as described above.  One can think about a finite temperature configuration in $0+1$ dimensions as either an explicit breaking of the time reparametrization symmetry or as being an anomaly, with the temperature as the background external field. Indeed, a ``central charge" appears in a number of our results.

We examine the thermodynamics associated with the $AdS_2$ plus aether solutions in Section IV.  These are characterized by the presence of a universal horizon, a surface beyond which even signals of arbitrary speed cannot reach infinity. The universal horizon therefore serves as a notion of causal boundary in a non-relativistic theory of gravity. We derive a thermodynamical relation of the form ${\cal E}  \sim T$ where ${\cal E}$ is the Noether charge associated with the global timelike Killing vector field and $T$ is the temperature of the universal horizon. Using the thermodynamic relations, we find there is an entropy associated with the universal horizon $S \sim \ln(\frac{T}{\Lambda})$, where $\Lambda$ is a new cutoff scale. We discuss the puzzling aspects and possible interpretations of this result.

Finally, motivated by the violation of the time reparametrization Ward identity, in Section V we study the algebra of charges associated with the asymptotic symmetries, following Brown and Henneaux \cite{Brown:1986nw}. Since the boundary geometry is one-dimensional (only time direction), conserved charges are simply evaluated at points and there is no integration over space. This causes problems when defining the Poisson bracket and a potential central charge independent of where it is evaluated on the boundary. If we instead make the ansatz that the charges are defined in terms of an integral over time, we find the potential central charge vanishes. This is at least consistent with the lore that there is no conformal anomaly in one-dimension. We interpret the violation of the Ward identity as being due to a novel type of explicit breaking of the time reparametrizations, caused by the presence of the aether. However, perhaps the ``central charge" appearing $AdS_2$ aether system can be seen as an artifact of the null dimensional reduction of an $AdS_3/CFT_2$ system.

\section{Einstein-aether theory in two dimensions}

Einstein-aether theory is a theory of gravity where the metric $g_{AB}$ is coupled to a dynamical unit timelike (co-)vector field $u_A$ \cite{AEtheory}. The aether field acts as a preferred frame at every point in spacetime, breaking local Lorentz invariance. To construct the Lagrangian
$L_{ae}(g_{AB}, u_A)$, one works in effective theory and writes down all possible terms up to second order in an expansion in derivatives of the metric and the aether. The result in four-dimensions is
\begin{align}
S_{ae} = \frac{1}{16\pi G_{ae}} \int d^{4} x \sqrt{-g} L_{ae} \label{totalaction} \ ,
\end{align}
where $L_{ae} = R + L_{vec} $,  with
\begin{align}
-L_{vec} = K^{AB}{}_{CD} \nabla_A u^C \nabla_B u^D - \lambda(u^2+1) \ ,
\end{align}
and
\begin{align}
K^{AB}{}_{CD} = c_1 g^{AB} g_{CD} + c_2 \delta^A_C \delta^B_D + c_3 \delta^A_D \delta^B_C - c_4 u^A u^B g_{CD} \ .
\end{align}
The coupling constants $c_i$ are dimensionless.

In \cite{Eling:2006xg}, Einstein-aether theory in two-dimensions was considered. In this lower-dimensional setting it was shown that the action reduces to the following form
\begin{align}
S_{ae} =  \int d^2 x \sqrt{-g} \left(  \frac{1}{2} \alpha F_{AB} F^{AB} + \beta (\nabla_A u^A)^2  + \lambda (u^2 + 1)  \right), \label{2daction}
\end{align}
where $F_{AB} = \nabla_A u_B - \nabla_B u_A$. In terms of the original $c_i$ coupling constants above, $\alpha = c_1 + c_4$ and $\beta = c_1+c_2 +c_3$. Also note that in two dimensions the Einstein-Hilbert term
leads to trivial dynamics, since the Ricci scalar is a total derivative. Finally, since the aether is twist-free and hypersurface orthogonal in two-dimensions, it defines a preferred time slicing.
Therefore two-dimensional Einstein-aether theory is equivalent to two-dimensional Ho\v{r}ava-Lifshitz gravity  \cite{Sotiriou:2011dr} .

Variation of the Lagrangian with respect to $g_{AB}$ and $u_A$ produces the metric equation of motion
\begin{align}
\alpha F_{AC} F_B{}^C - \frac{1}{2} g_{AB} \left(\frac{1}{2} \alpha F^{CD} F_{CD} - \beta (\nabla_C u^C)^2 - 2 \beta u^C \nabla_C (\nabla_D u^D) \right) - 2 \beta u_{(A} \nabla_{B)} \nabla_C u^C + \lambda u_A u_B = 0
\end{align}
and the aether field equation
\begin{align}
\alpha \nabla_B F^{BA} + \beta \nabla^A (\nabla_C u^C) - \lambda u^A = 0.
\end{align}
The Lagrange multiplier $\lambda$ can be found by multiplying the aether field equation with $u_A$ and using the unit constraint. The solutions to these field equations were found and analyzed in \cite{Eling:2006xg}. In particular, when
$\alpha = \beta$ there are only flat spacetime solutions. When $\alpha \neq \beta$ there are non-constant and constant curvature solutions. In the second class an $AdS_2$ solution with an aether field was found. In Fefferman-Graham
like coordinates for the Poincare patch, one finds the solution
\begin{align}
ds^2 = -r^2 dt^2 + \frac{dr^2}{r^2} \nonumber \\
u_A dx^A = k r dt - \frac{\sqrt{k^2 - 1}}{r}  dr ,  \label{vacuum}
\end{align}
where $k = \sqrt{(\beta-\alpha) \beta}/(\alpha-\beta)$. We take $\alpha$ and $\beta$ to be positive and $\beta > \alpha$. Note that no cosmological constant term is needed for this configuration to be a solution\footnote{Note that we can include a cosmological constant term $\Lambda$ in the two-dimensional Einstein-aether action. However, this only affects the solution (\ref{vacuum}) by changing the value of $k$}. A plot of the flow lines of the aether for this solution on the Penrose diagram of AdS can be found in Figure 4 of \cite{Eling:2006xg}. The aether field is regular in the Poincare patch, but becomes singular on the Poincare horizon. In two-dimensions, the boundary of $AdS_2$ is disconnected into two separate boundaries. From the holographic point of view this raises the question of whether the dual description is terms of a single $CFT_1$ or two systems on the boundaries. In this paper we will consider the theory in the Poincare and smaller sub-patches of the spacetime, which appears to restrict us to only one boundary system.

\section{Asymptotic Symmetry Group}

To investigate the potential holographic dual to this solution, we will analyze the asymptotic symmetries, in the spirit of Brown and Henneaux. To start, we consider the solution in (\ref{vacuum}).
We want to find an asymptotic Killing vector, i.e. a Killing vector $\xi^A$ that preserves the following asymptotically AdS boundary conditions
\begin{align}
g_{tt} = -r^2 + O(1), ~~ g_{tr} = O(1/r^3), ~~ g_{rr} = O(1/r^4) \nonumber \\
u_t = k r + O(1), ~~ u_r = -(\sqrt{k^2-1})/r + O(1/r^2)
\end{align}
The result is
\begin{align}
\xi^t = \epsilon(t) + \frac{1}{2} \frac{1}{r^2} \partial^2_t \epsilon + O(1/r^4) \nonumber \\
\xi^r = -r \partial_t \epsilon + O(1/r^2) \label{asympvector}
\end{align}
for arbitrary function $\epsilon(t)$, associated with an infinitesimal $t \rightarrow t + \epsilon(t)$. This is exactly the asymptotic Killing vector that arises in studies of asymptotic symmetries in pure $AdS_2$, see e.g. \cite{Hotta:1998iq} . The aether field does not explicitly break the asymptotic symmetry group, which is the infinite dimensional set of one-dimensional conformal transformations. These can be thought of as ``one-half" of the conformal transformations in two-dimensions, which lead to the Virasoro algebra. Here any mapping $t \rightarrow f(t)$ takes the metric $ds^2 = -dt^2$ into $ds^2 = -f'(t)^2 dt^2$.

We now parametrize the first order corrections to the metric and aether in the following way
\begin{align}
g_{tt} = -r^2 + s_{tt} + \cdots \nonumber \\
u_t = k r +\sqrt{k^2-1} \rho_t, ~~ u_r = -\frac{\sqrt{k^2-1}}{r} - \frac{k \rho_t}{r^2} + \cdots   \label{perturbedsoln}
\end{align}
Under infinitesimal diffeomorphisms generated by (\ref{asympvector}) one finds
\begin{align}
\delta_{\xi} s_{tt} =   2  s_{tt} \partial_t \epsilon + \epsilon \partial_t s_{tt} - \partial^3_t \epsilon \nonumber \\
\delta_{\xi} \rho_t  = \epsilon \partial_t \rho_t +  \rho_t \partial_t \epsilon +  \partial^2_t \epsilon .\label{changes}
\end{align}
Asymptotic symmetries are always spontaneously broken. For example, consider the case where $s_{tt} = \rho_{t} = 0$, which corresponds to a choice of vacuum state.   This configuration is only invariant under transformations $\epsilon = (1, t)$, which correspond to infinitesimal time translations and an overall scale transformation. This affine subgroup $A(1)$ is isomorphic to the Lorentz subgroup of boosts and null rotations (Lorentz transformations preserving null vectors) in three-dimensional Minkowski spacetime. These are the exact symmetries of the metric and aether configuration. Thus there is a spontaneous breaking of time reparametrizations down to $A(1)$.  Usually the $AdS_2$ vacuum is invariant also under infinitesimal special conformal transformations generated by $ \epsilon(t) = t^2$, and one has the $SL(2)$ symmetries, but the presence of the aether breaks this down to $A(1)$. In the field theory we could interpret this as the usual $SL(2)$ invariant vacuum state plus a source associated with the aether field.

\section{Thermodynamics and Conserved Charges}

Now suppose we consider the case of a finite diffeomorphism of (\ref{vacuum}) preserving the gauge and boundary conditions. One finds
\begin{align}
\delta_{\xi} s_{tt} = - 2 \{f,t\} \nonumber \\
\delta_{\xi} \rho_t = \frac{\ddot{f}}{\dot{f}},
\end{align}
where $\{f,t\}$ is the Schwarzian derivative
\begin{align}
\{f,t\}  = \frac{\dddot{f}(t)}{\dot{f}(t)}- \frac{3}{2}\frac{\ddot{f}(t)^2}{\dot{f}(t)^2},  \label{Schwarzian}
\end{align}
and the dot represents a time derivative. Taking, for example, $f(t) = e^{r_0 t}$ one can express the metric to all orders in $1/r$ as
\begin{align}
ds^2 &= -(r^2-r_0^2) d\tau^2 + \frac{dr^2}{r^2-r_0^2} \nonumber \\
u_A dx^A &= (k r + \sqrt{k^2-1} r_0) d\tau + \left(\frac{k r + \sqrt{k^2-1} r_0 + \sqrt{k^2-1} r + k r_0}{r_0^2-r^2}\right) dr, \label{BHSoln}
\end{align}
which is the $AdS_2$ black hole (or AdS-Rindler coordinates) plus the aether configuration. One can verify that this is indeed a solution to the field equations.

We can also express the above metric in a Ho\v{r}ava-Lifshitz gauge associated with the time foliation (slices of constant $u$)
\begin{align}
ds^2 = -(r^2 - r_0^2) du^2 + 2 N_r du dr + \frac{1-N_r^2}{r^2-r_0^2} dr^2 \nonumber \\
u_A dx^A = (kr + r_0 \sqrt{k^2-1}) du,
\end{align}
where
\begin{align}
N_r = \frac{r \sqrt{k^2-1} + k r_0}{k r + r_0 \sqrt{k^2-1}}
\end{align}
is the shift vector. From this form, we see that there is a universal horizon, defined as the location where the dot product of the global timelike Killing vector $\chi^A = (\partial/\partial t)^A$ with $u_A$ vanishes
\begin{align}
r_{UH} = -\frac{\sqrt{k^2-1}}{k} r_0 = \sqrt{\frac{\alpha}{\beta}} r_0 .
\end{align}
The region beyond this horizon is causally disconnected from infinity, even for signals of arbitrary speed and therefore defines a notion of black hole. In \cite{Berglund:2012fk,Janiszewski:2014iaa} it has been argued there is a Hawking temperature associated with universal horizons, which has the form
\begin{align}
T_{UH}  = \left(\frac{a^A s_A |\chi |}{2\pi}\right)_{r=r_{UH}},
\end{align}
where $a^A = u^B \nabla_B u^A$ and $s^A$ is the unit vector orthogonal to $u^A$.\footnote{In \cite{Janiszewski:2014iaa} it was argued that the Hawking temperature in \cite{Berglund:2012fk}, which was obtained by the tunnelling method, is off by a factor of two. Here we will use the form in \cite{Janiszewski:2014iaa}.} Evaluating this formula for our solution, we find
\begin{align}
T_{UH} = \frac{r_0}{2\pi}.   \label{TUH}
\end{align}
Note that this is consistent with the exponential relation between the Poincare time $t$ and the Schwarzschild-like time $\tau$. There is a periodicity in imaginary time with period $\beta = 2\pi/r_0$. This indicates a potential dual configuration at the boundary is at finite temperature.

One important question is the nature of the conserved charges corresponding to the asymptotic Killing vectors. One way to  extract these charges is to employ the covariant phase space approach of Wald \cite{Wald:1993nt}. In general, the variation of the Lagrangian density $L$ leads to
\begin{align}
\delta L = E_i \delta \psi^i + \nabla_A \theta^A,
\end{align}
where $\psi^i$ are the fields in the problem, $E_i$ are the equations of motion, and $\theta^A(\psi^i, \delta \psi^i)$ is the symplectic potential current density. By acting on this equation with two variations, one can show that on-shell
\begin{align}
\nabla_A \omega^A = 0,
\end{align}
where $\omega^A = \delta_1 \theta^A(\delta_2) - \delta_2 \theta^A(\delta_1)$. The symplectic form $\omega$ is defined as the integral over a Cauchy slice
\begin{align}
\omega = \int_{\Sigma} d\Sigma_A \omega^A.
\end{align}
For diffeomorphisms generated by a vector field $\xi^A$, the field variations are Lie derivatives. From Hamilton's equations of motion, the variation of the Hamiltonian associated with $\xi^A$ is
\begin{align}
\delta {\cal H}_{\xi}  = \int_{\Sigma} d\Sigma_A \omega^A (\psi^i, {\cal L}_\xi \psi^i).
\end{align}
This equation can be expressed in first in terms of the Noether current density $J^A = \theta^A(\psi^i, {\cal L}_\xi \psi^i) - \xi^A L$,
\begin{align}
\delta {\cal H}_{\xi}  = \int_{\Sigma} d\Sigma_A  \left( \delta J^A - 2 \nabla_B (\theta^{[A} \xi^{B]}) \right),
\end{align}
and finally in terms of a surface integral and the antisymmetric Noether potential density $Q^{AB}$
\begin{align}
\delta {\cal H}_{\xi} = \int_{\partial \Sigma} d n_{AB} \left( \delta Q^{AB} - \theta^{[A} \xi^{B]} \right),
\end{align}
where $J^A = 2 \nabla_B Q^{AB}$. The surface element $n_{AB}$ is $2 r_{[A} t_{B]}$, where $r_A$ and $t_A$ are the unit norms to a surfaces of constant $r$ and $t$ respectively. A Hamiltonian exists for the asymptotic Killing vectors if there is a $B^A$ such that $\delta \int_{\infty} dn_{AB} B^{[A} \xi^{B]} = \int_{\infty} dn_{AB} \theta^{[A} \xi^{B]}$.

For Einstein-aether theory, the form of the symplectic current and Noether potentials was found generally in \cite{Foster:2005fr,Mohd:2013zca}. In the two-dimensional case we find
\begin{align}
\theta^A = \sqrt{-g} \left(\beta (\nabla_C u^C) (u^A g^{BC} - 2 g^{AB} u^c) \delta g_{BC} + 2\alpha F^{AB} \delta u_B + 2 \beta (\nabla_C u^C) g^{AB} \delta u_B \right)
\end{align}
and
\begin{align}
Q^{AB} = -\sqrt{-g} \left(\alpha F^{AB} (u_C \xi^C) + \beta (\nabla_C u^C) (u^A \xi^B - u^B \xi^A) \right).
\end{align}
Computing the Hamiltonian associated with the asymptotic Killing vector (\ref{asympvector}), yields
\begin{align}
{\cal H}_{\xi} = 2 \sqrt{\alpha \beta} \left( \epsilon \rho_t  + \dot{\epsilon} \right)  \label{charge}
\end{align}
The only contribution to the integral at infinity (here just an evaluation at the boundary) comes from the Noether current density. The last term can be thought of as an integration constant since it does not depend on the variation of the fields. One can re-define the ${\cal H}$ by a shift such that for the background configuration where $\rho_t = 0$ it vanishes, i.e.
\begin{align}
{\cal H'}_\xi = 2 \sqrt{\alpha \beta}~ \epsilon \rho_t.   \label{charge2}
\end{align}
We will work with this form from this point forward.

Another useful way to compute the charge is via the holographic (Brown-York) stress tensor. Here we consider the on-shell gravitational action, which is a boundary term. For Einstein-aether theory, the effective action should depend on the boundary metric $\gamma_{\mu \nu}$ and boundary aether $v_{\mu}$.
The variation of the effective action $W(\gamma,v)$ can be expressed as
\begin{align}
\delta W = \int d^d x \sqrt{\gamma} \left(\frac{1}{2} E^{\mu \nu} \delta \gamma_{\mu \nu} + J^\mu \delta v_\mu \right),   \label{variation}
\end{align}
where $E^{\mu \nu} = \frac{2}{\sqrt{\gamma}} \frac{\delta W}{\delta \gamma_{\mu \nu}}$ and $J^\mu = \frac{1}{\sqrt{\gamma}} \frac{\delta W}{\delta v_{\mu}}$. Demanding diffeomorphism invariance of the action $W(\gamma,v)$, one finds the following Ward identity
\begin{align}
\delta_\xi W = 0 = \int d^d x \sqrt{\gamma} \left( E^{\mu \nu} D_{\mu} \xi_\nu + J^\mu {\cal L}_\xi v_\mu \right),
\end{align}
where $D_\mu$ is the covariant derivative associated with the metric $\gamma_{\mu \nu}$. We can express this equation as
\begin{align}
D_\mu (E^{\mu \nu} + J^\mu v^\nu) = -J^\mu D^\nu v_\mu . \label{stressWard}
\end{align}
In the following we will take the natural definition of the stress tensor to be
\begin{align}
T^{\mu \nu} = E^{\mu \nu} + J^\mu v^\nu,
\end{align}
Note that this form is equivalent to the (non-symmetric) stress tensor that is obtained via a variation of the vielbein instead of the metric as the fundamental field (see, e.g. \cite{Arav:2016xjc}). The Ward identity associated with one-dimensional conformal transformations
yields
\begin{align}
\delta_\sigma W = 0 = \int d^d x \sigma \left( E^\mu_\mu + J^\mu v_\mu \right),
\end{align}
which seems to imply, in one-dimension, a vanishing energy $T^{tt} = 0$. The associated charge\footnote{This charge has the same value on any surface of constant time since the contribution from the right hand side of (\ref{stressWard}) vanishes at infinity.} is
\begin{align}
{\cal H} = \int_{\partial \Sigma}  T_{\mu \nu} \xi^\mu d \Sigma^\nu.
\end{align}

To compute the stress tensor, we vary the bulk Einstein-aether action and impose the field equations. The result is
\begin{align}
\delta W = \delta S_{ae} = \int dt \sqrt{h} \left[ \beta \left((r_C u^C) h^{AB} - 2 r^{(A} u^{B)} \right) \nabla_C u^C \delta g_{AB} + \right. \nonumber \\  \left.
 \left( 2 \alpha (r^B \nabla_B u^A - g^{AB} r^C \nabla_B u_c) + 2 \beta (\nabla_B u^B) r^A \right) \delta u_A \right]
\end{align}
where $h_{AB} = g_{AB} - r_A r_B$. In two-dimensions the only non-zero part of the stress tensor is $T^{tt}$. Using $\delta g_{AB} = r^2 \delta \gamma_{tt} + \cdots$, $\delta u_A = r \delta v_t + \cdots$ and $h_{tt} = r^2 \gamma_{tt}$, we can extract from this expression
$E^{tt}$ and $J^t$ and find the value for $T^{tt}$ for the metric (\ref{perturbedsoln}) in the limit as $r \rightarrow \infty$. The final result agrees with (\ref{charge2}).

In the case where $\epsilon_1(t) = 1$, the asymptotic Killing vector is a global symmetry and the corresponding charge corresponds to an energy of the system. If we use the Hawking temperature at the universal horizon (\ref{TUH}), we find the thermodynamic
relation
\begin{align}
{\cal E} = 2 \sqrt{\alpha \beta} ~r_0 = 4 \pi \sqrt{\alpha \beta}~ T.
\end{align}
Note that this formula is similar to the one found for two-dimensional CFT's at finite temperature, $P = 4 \pi^2 c_{2d} T^2$, relating pressure to central charge \cite{Bloete:1986qm,Affleck:1986bv}. One considers a conformal transformation that maps the plane into the cylinder. Using the formula for the transformation of the stress tensor under a conformal mapping, one can show that the vacuum acquires an energy. This can be interpreted as a Casimir energy since the system now has an effective finite size. Here, if we act with an asymptotic diffeomorphism, we find from (\ref{changes})
\begin{align}
\delta_\xi {\cal E} =  \dot{f} {\cal E} + 2 \sqrt{\alpha \beta} \frac{\ddot{f}}{\dot{f}}
\end{align}
The first term has the form of the transformation of a vector current under diffeomorphism, while the last is an anomalous term, with a ``central charge" of $2 \sqrt{\alpha \beta}$.  When we start from the ${\cal E}=0$ vacuum and set $f = e^{r_0 t}$, this yields the above result. Thus the energy here can be thought of as a Casimir energy arising from the mapping of the system from the line to a circle.

To obtain the entropy of the system we use the thermodynamic relation
\begin{align}
S = \int \frac{d {\cal E}}{T'} = \int^{T}_{\Lambda} \frac{1}{T'} \frac{d{\cal E}}{dT'}~  dT'.
\end{align}
Inserting the formula for the energy, we find
\begin{align}
S = 4\pi \sqrt{\alpha \beta} \ln \left(\frac{T}{\Lambda}\right),
\end{align}
where $\Lambda$ is a new ``spontaneously generated" scale  (integration constant). It acts as a cutoff since formally the number of states in the system is infinite. One can also extract the free energy of the system using $F = {\cal E} - TS$. This yields
\begin{align}
F = 4 \pi \sqrt{\alpha \beta} T \left(1 - \ln \left(\frac{T}{\Lambda}\right) \right).
\end{align}

In Einstein-aether theory there is no general Wald formula for the entropy but in spherically symmetric black holes in higher dimensions it has been argued the entropy is proportional to the area of the universal horizon \cite{Mohd:2013zca,Janiszewski:2014iaa}. In two-dimensions though the horizon is a point. Therefore our result is a new prediction for horizon entropy in two-dimensional Einstein-aether theory.

A logarithmic dependence of entropy on temperature has been found previously in the Almheiri-Polchinski dilaton model \cite{Almheiri:2014cka}, where it is the contribution at one-loop to the thermodynamical entropy from (conformal) scalar matter fields.  Spradlin and Strominger also found a logarithmic dependence on temperature in the entanglement entropy for conformal scalar fields outside an $AdS_2$ black hole \cite{Spradlin:1999bn}\footnote{Note that $\ln(T/\Lambda)$ acts like the dilaton for the spontaneous breaking of conformal symmetry by finite temperature \cite{Eling:2013bj}. Perhaps here, where such a spontaneous breaking is explicit, this factor does indeed measure the number of states.}. The formulas for two-dimensional entropy agree if we identify the factor $4\pi \sqrt{\alpha \beta}$ as, again, being proportional to a central charge. However, a direct connection with these past results, which are obtained at one-loop, is not clear. It may be that one can consider the aether field as a type of matter field on the $AdS_2$ background and the entropy is an entanglement entropy associated with that field.

There are in principle two puzzling features to the logarithmic dependence.  For $\Lambda > T$ the entropy is negative, and as $T \rightarrow 0$ the entropy $S \rightarrow - \infty$. The zero temperature state is of course the original Poincare vacuum (\ref{vacuum}). One could argue that $\Lambda < T$ and that as $T \rightarrow 0$, we should also effectively take the cutoff $\Lambda \rightarrow 0$, such that the entropy in the vacuum state vanishes. Essentially $T \sim \Lambda$ is where the theory is strongly coupled and the semi-classical picture of Hawking radiation breaks down. However, if we were to take negative entropy seriously, in quantum information theory there is a notion of a conditional entropy $H({\cal S}| O)$, which can be negative and has a thermodynamic interpretation \cite{nature1,nature2}. This entropy depends on the amount of information an observer $O$ has about some quantum system ${\cal S}$. One could imagine that the entropy associated with the universal horizon is a measure of the ignorance of an observer in the preferred frame about the dual quantum system. Note that in the case of the Poincare vacuum, the universal horizon coincides with the extremal Killing horizon. Here the aether field becomes singular and infinitely stretched, which could be linked to the divergence of the entropy.

\section{Algebra of Charges}

We now investigate whether the violation of the time reparametrization Ward identity could be associated with an anomaly. One way to determine if this is the case is to consider whether the algebra of the conserved charges actually has a central extension. We will first consider the bracket of two asymptotic Killing vectors, $[\xi_1, \xi_2]^A$. This is  defined as the Lie Bracket $\xi^B_1 \partial_B \xi^A_2 - \xi^B_2 \partial_B \xi^A_1$. In general one must consider a modified bracket \cite{Barnich:2001jy},that subtracts off potential changes in $\xi_2$
due to variations ${\cal L}_{\xi_1} g_{AB}$ or ${\cal L}_{\xi_1} u_A$ and visa versa. In this case, these charges are higher order ($O(r^{-4}$)), so the standard Lie bracket is suitable. One finds
\begin{align}
[\xi_1, \xi_2]^t = \epsilon_1 \dot{\epsilon}_2 - \epsilon_2 \dot{\epsilon}_1.
\end{align}
One typically expands the function $\epsilon(t)$ in terms of a basis of polynomials
\begin{align}
\epsilon(t) = -\sum^{\infty}_{m=-\infty} a_m t^{m+1}.  \label{expansion}
\end{align}
Note that the $m=-1$ corresponds to time translations, $m=0$ to scale transformations, and $m=1$ to special conformal transformations. Denoting $\epsilon_m = -a_m t^{m+1}$, one can show as usual that the Lie bracket leads to the Witt algebra
\begin{align}
[\epsilon_m, \epsilon_n] = (m-n) \epsilon_{m+n}
\end{align}
associated with one-dimensional diffeomorphisms. In this one-dimensional case we only have one copy of the Witt algebra, instead of the two copies in two-dimensions. The generators $(-1,0,1)$ form a sub-algebra since for these cases the vector fields are finite at zero and infinity. However, in this case, as we noted earlier, one has to be careful because the generator of special conformal transformations is \textit{not} an exact symmetry of the system.

Following original work of Brown and Henneaux, which has been elaborated on in for example, \cite{Koga:2001vq, Barnich:2001jy,Silva:2002jq,Barnich:2010eb}, one can show that the conserved Noether charges associated with the asymptotic vectors satisfy the following algebra
\begin{align}
\Big[{\cal H}_{\xi_1}(g,u), {\cal H}_{\xi_2}(g,u)\Big]_P = {\cal H}_{[\xi_1, \xi_2]}(g,u) + K_{\xi_1, \xi_2}  \label{chargealgebra}
\end{align}
The bracket on the left hand side represents the Poisson (or Dirac) bracket of the conserved charges. The term $K_{\xi_1, \xi_2}$ does not depend on the dynamical fields and therefore acts as a central
term in the algebra. The Poisson bracket for the charges has been typically defined as
\begin{align}
\Big[{\cal H}_{\xi_1}(g,u), {\cal H}_{\xi_2}(g,u)\Big]_P = \delta_{\xi_2} {\cal H}_{\xi_1}
\end{align}
where $\delta_{\xi_2} \xi_1 = 0$, meaning that the variation acts only on the fields. As a result,
\begin{align}
 \delta_{\xi_2} {\cal H}_{\xi_1} = 2\sqrt{\alpha \beta} \left( \epsilon_1 \dot{\epsilon}_2 \rho_t + \epsilon_1 \epsilon_2 \dot{\rho}_t + \epsilon_1 \ddot{\epsilon_2} \right)
\end{align}
On the other hand,
\begin{align}
{\cal H}_{[\xi_1,\xi_2]} = 2\sqrt{\alpha \beta} (\epsilon_1 \dot{\epsilon}_2 - \epsilon_2 \dot{\epsilon}_1) \rho_t.
\end{align}
These results do not appear to be consistent with (\ref{chargealgebra}). In, for example, the $AdS_3$ and BMS cases, one can show that (\ref{chargealgebra}) holds by evaluating $\delta_{\xi_2} Q_{\xi_1}$ and integrating by parts over spatial direction. In those cases the total charges were integrals over space. This is not the case in one-dimension where no spatial integrals are present and one is evaluating at a point on the boundary. One should also have an antisymmetry $\delta_{\xi_2} Q_{\xi_1} = - \delta_{\xi_1} Q_{\xi_2}$, which is not obviously true above.  One way to proceed is to define the one-dimensional Poisson bracket so that antisymmetry is made manifest
\begin{align}
\Big[{\cal H}_{\xi_1}(g,u), {\cal H}_{\xi_2}(g,u)\Big]_P = \left( \delta_{\xi_2} {\cal H}_{\xi_1} - \delta_{\xi_1} {\cal H}_{\xi_2} \right). \label{Poisson2}
\end{align}
Then (\ref{chargealgebra}) does hold and one finds the central-like term
\begin{align}
K_{\xi_1, \xi_2} = 2 \sqrt{\alpha \beta} \left(\epsilon_1 \ddot{\epsilon}_2 - \epsilon_2 \ddot{\epsilon}_1 \right).
\end{align}
However, note that this expression depends on time in that we must evaluate it at some $t=t_0$. Again, comparing to the $AdS_3$ case, the discrepancy is due to the lack of a spatial integral. If we expand the analogous $AdS_3$ expression into modes $e^{i m(t \pm \phi)}$ and integrate over $\int^{2\pi}_{0} d \phi$, one finds that the non-vanishing piece of the central term is independent of time.

A possible resolution is to define a total time independent charge in terms of an integral over time (and invoking a periodicity in imaginary time)
\begin{align}
H_\xi = 2 \sqrt{\alpha \beta} \int^{2\pi}_{0} dt \epsilon(t) \rho_t
\end{align}
Then if (\ref{Poisson2}) holds for $H_\xi$, we find
\begin{align}
K_{\xi_1, \xi_2} = 2 \sqrt{\alpha \beta} \int^{2\pi}_{0} dt  \left(\epsilon_1 \ddot{\epsilon}_2 - \epsilon_2 \ddot{\epsilon}_1 \right).
\end{align}
If we expand $\epsilon(t)$ into Fourier modes $e^{i m t}$, for integer $m$ and $n$, we find that $K_{m,n}$ vanishes for all $(m,n)$. This indicates this potential charge algebra is without a central term.

\section{Discussion}

It is difficult to interpret the non-zero energy via an anomaly in the one-dimensional conformal symmetry.  Therefore we instead interpret the violation of the Ward identity as a type of explicit breaking of the time reparametrization symmetry. A finite temperature is a soft breaking, introducing an effective length scale in $T$. However, in one dimension scale invariance implies that the density of states must scale like $\rho({\cal E}) = A \delta({\cal E}) + B/{\cal E}$ \cite{Jensen:2016pah}. The first term is a possible zero temperature entropy, while the second is the $T^{-1}$ term we found from the black hole thermodynamics. This leads to the presence of the logarithm in the entropy and free energy and means there must be another cutoff scale $\Lambda$ generated as well. Thus we have a ``spontaneous explicit breaking" supported by the presence of the aether. It would be interesting to understand a potential holographic dual in more detail. Our results may also be useful for the study of various condensed matter systems via $AdS_2$ holography, e.g. \cite{Liu:2009dm,Faulkner:2009wj}.

Finally, it is possible that two-dimensional Einstein-aether theory can be realized as a dimensional reduction of a gravity theory in $AdS_3$, along the lines of the Einstein-Maxwell-dilaton models discussed in \cite{Castro:2008ms,Cvetic:2016eiv}. For example, it is known that non-relativistic theories are the result of a null reduction of gravity on higher dimensional Lorentzian manifolds \cite{Julia:1994bs,Jensen:2014aia}. Perhaps the central charge and logarithmic scaling found here have their origins in the non-relativistic limit of a two-dimensional CFT.

\section*{Acknowledgments}

I would like to thank T. Andrade, Y. Oz, and A. Starinets for valuable discussions. This research was supported by the European Research Council
under the European Union's Seventh Framework Programme (ERC Grant agreement 307955).

\end{document}